\documentclass{icrc2009}

\usepackage{graphicx}   
\usepackage{caption}    
\usepackage[font=footnotesize]{subfig} 
\usepackage{fixltx2e}
\usepackage{url}

\newcommand{\shorttitle}[1]%
{\markboth{Proceedings of the 31\MakeLowercase{$^{st}$} ICRC, {\L}\'{o}d\'{z} 2009}{#1} }
\newcommand{\etal}{\MakeLowercase{\textit{et al. }}} 

\begin{document}
\title{A study of the Correlation of Arrival Directions of UHECRs with
the Large Scale Structure of the Universe }

\author{\IEEEauthorblockN{Dongsu Ryu\IEEEauthorrefmark{1},
			  Hyesung Kang\IEEEauthorrefmark{2}, and
                          Santabrata Das\IEEEauthorrefmark{3}}\\

\IEEEauthorblockA{\IEEEauthorrefmark{1}Department of Astronomy \& Space
          Science, Chungnam National University, Daejeon 305-764, Korea}
\IEEEauthorblockA{\IEEEauthorrefmark{2}Department of Earth Sciences, Pusan
          National University, Pusan 609-735, Korea}
\IEEEauthorblockA{\IEEEauthorrefmark{3}Korea Astronomy and Space Science
          Institute 61-1, Hwaam Dong, Yuseong-Gu,Daejeon 305 348, Korea}}

\shorttitle{Ryu \etal Correlation Study of UHECR Arrival Directions}
\maketitle

\begin{abstract}
Ultrahigh energy cosmic rays (UHECRs) are believed to originate from
astrophysical sources, which should trace the large scale structure
(LSS) of the universe.
On the other hand, the magnetic field in the intergalactic space
(IGMF), which also traces the LSS of the universe, deflects the
trajectories of the charged UHECRs and spoils the positional correlation
of the observed UHECR events with their true sources.
To explore this problem, we studied a simulation of the propagation of
UHE protons through the magnetized LSS of the universe, reported
earlier in Das et al. (2008), in which the IGMF was estimated based
on a turbulence dynamo model (Ryu et al. 2008).
Hypothetical sources were placed inside clusters and groups of galaxies
in the simulated universe, while observers were located inside groups of
galaxies that have similar properties as the Local Group.
We calculated the statistics of the angular distance between the arrival
directions of simulated UHE proton events and the positions of candidate
sources in our simulation. 
We compared the statistics from our simulation with those calculated
with the Auger data.
We discussed the implication of our works on the nature of the sources
of UHECRs.
\end{abstract}

 
\section{Introduction}

Recently the Pierre Auger Collaboration reported a significant 
anisotropy in the arrival directions of ultrahigh energy cosmic rays
(UHECRs) \cite{augerScience}.
In their study, the nearest neighbor of a given UHECR event is identified
among the active galactic nuclei (AGNs) listed in the 12th edition of
Veron-Cetty (VC) catalog \cite{vcv06}. 
The maximum correlation was observed for 20 events out of total 27 events
with energy above $57$ EeV, when the closest AGN is searched within the
angular separation of $3.2^{\circ}$ among 442 AGNs with redshift
$z \le 0.018$ (corresponding to the distance $D \le 75$ Mpc for
$h^{-1} = 0.7$). 
Their study demonstrated a potential association of UHECR events with
the sky position of nearby AGNs. 
However, the statistical significance of this correlation analysis may
not yet be robust, since the number of detected events at extremely
high energy is still rather small. 
But at least it implies that the sources of UHECRs might be correlated 
with the matter distribution in the local universe.

In the correlation analysis, a priori knowledge of the intergalactic
magnetic field (IGMF) is essential as it guides the propagation of UHECRs
through the intergalactic space.
Even with considerable observational efforts, the nature of the IGMF is
still poorly known, especially in the low-density regions like filaments,
sheets and voids, where activities are relatively weak. 
In Ryu et al. (2008) \cite{ryuetal08}, we suggested that a part of
the gravitational energy is transferred to the magnetic field energy as a
result of the turbulent dynamo amplification of weak seed fields in the
large scale structure (LSS) of the universe.
This model predicts that $\sim 1$ \% of the intergalactic space is
magnetized with the field strength $B > 10$ nG, and that the IGMF follows
largely the matter distribution in the cosmic web. 

Adopting this IGMF model, we estimated that only $\sim 35$ \% of UHE
protons above 60 EeV may arrive within 5$^{\circ}$ of their sources
located inside the GZK sphere with radius of 100 Mpc \cite{dasetal08}. 
Considering that the mean deflection angle of super-GZK protons caused
by the IGMF in our model is $\sim 15^{\circ}$ \cite{dasetal08}, it is
natural to expect that, in the correlation study of the Auger UHECR
events, a significant fraction of the identified source candidates
within 3.2$^{\circ}$ may not be the true sources of the UHECRs.
In order to explore how the distribution of source candidates and the
IGMF in the local universe influence the correlation analysis of UHECRs,
we used the simulation results reported earlier in Das et al (2008)
\cite{dasetal08}.

\section{Simulation Setup}

Following \cite{augerScience}, we set the maximum limit of the
source-to-observer distance as $75$ Mpc.
Inside the simulation volume of radius 75 Mpc, $\sim 500$ source candidates
(comparable to the number of AGNs from VC catalog) were placed in the
highest temperature regions of the cosmic web.
Observers were placed in the regions corresponding to groups of galaxies with
the halo temperature similar to that of our Local Group.
For each UHE proton event above 60 EeV, we computed two angular distances:
{\it separation angle} and {\it deflection angle}.
The separation angle, $S$, is the angular distance between the arrival
direction of a given event and the position of the nearest AGN among 500
source candidates. 
The deflection angle, $\theta$, is the angular distance between the arrival
direction of a given event and its true source position. 
Note that the nearest AGN may be not the true source of that event, and
$S$ would be on average smaller than $\theta$.
Here, we denote $D_S$ as the distance to the nearest AGN with the separation
angle $S$, and $D_{\theta}$ as the distance to the true source with the
deflection angle $\theta$. 

In our simulated universe, the intergalactic space was magnetized by
turbulent plasma motions generated during the structure formation
\cite{ryuetal08}.
For simplicity, the injection proton spectrum at source locations was
characterized by a power low in the energy range above $6 \times 10^{19}$
eV with the exponent of $\gamma=2.7$.

\section{Results}

\begin{figure}[!t]
\centering
\includegraphics[width=3.2in]{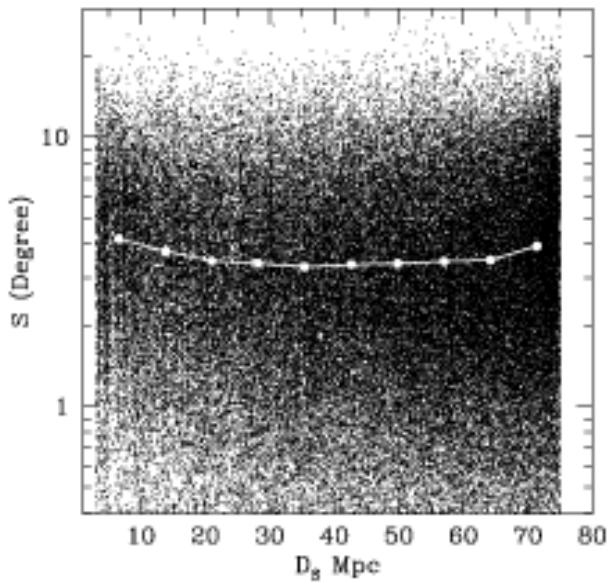}
\caption{
Distribution of separation angle ($S$) as functions of distance ($D_S$)
of the nearest AGNs in the sky inside a sphere of radius 75 Mpc. 
Black dots are for the events recorded in our simulation.
White circles connected with solid line represent the mean values,
$\langle S \rangle$, in the distance bin of $[D_S, D_S+d D_S]$.}
\end{figure}

\begin{figure}[!t]
\centering
\includegraphics[width=3.2in]{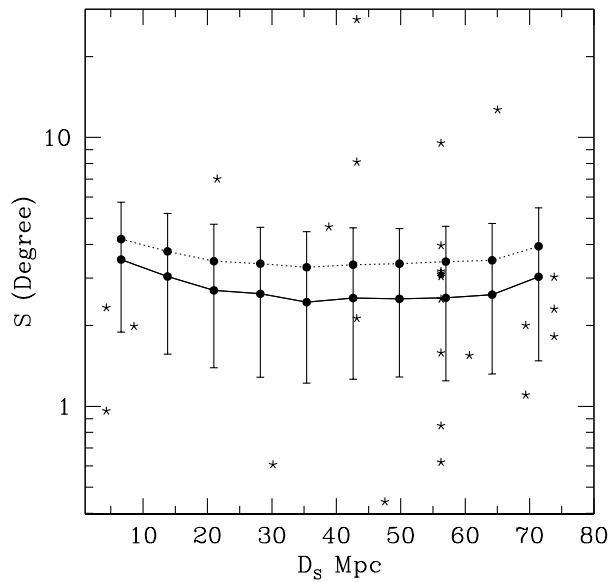}
\caption{Mean (dotted line) and median (solid line) values of separation
angle, $S$, as functions of $D_S$ for the recoded events in our simulation. 
The solid vertical lines with marks connect the first and third quartiles
in the given $D_S$ bin.
Asterisks denote the highest energy Auger events that are correlated with
the sky position of nearby AGNs \cite{auger08}.}
\end{figure}

\subsection{Separation Angle and Deflection Angle}

In Figure 1, we show the distribution of separation angle ($S$) versus
the distance of the nearest source ($D_S$).
Each dot denotes a recorded event obtained from simulation.
The white circles connected with solid line represent the mean values
of $S$ for the events from AGNs with the distance bin of $[D_S, D_S+d D_S]$.
We note that the mean separation angle $\langle S \rangle$ shows a U-shape
distribution (see the discussion below).
Overall $\langle S \rangle$ is smaller than $\sim 4^{\circ}$ for all $D_S$,
and does not change sensitively with $D_S$. 
The mean separation angle over all the simulated events is
$\langle S \rangle_{\rm sim} = 3.6^{\circ}$. 

In Figure 2, we compare the results of our simulation 
with the Auger data \cite{auger08}. 
Asterisks denote the highest energy 27 Auger events that are correlated
with the position of nearby AGNs listed in VC catalog.
The mean separation angles for the 20 events that gives the maximum
correlation, for the 26 events that excludes the event with $S> 27^{\circ}$
and for all the 27 events are 
$\langle S \rangle _{\rm Auger}$ = $1.91^{\circ}$, $3.23^{\circ}$ and
$4.13^{\circ}$, respectively.
The filled circles connected with dotted line represent the mean values of
$S$ (same as those in Figure 1), while the filled circles connected with
solid line refer to the median values of $S$ for the simulated events.  
For a given distance bin, $[D_S, D_S+d D_S]$, the solid vertical lines
with marks on the both sides of the median values (the first and third
quartiles) provide the measure of dispersion on $S$.
Note that 15 out of 27 Auger events lie within the quartile marks obtained
from our simulation data. 

Figure 3 shows the distribution of the deflection angle ($\theta$) of
UHE proton events with truce sources at distance $D_{\theta}$.
Again each dot represents a simulated event.
The white circles connected with dotted and solid lines are for the mean
and the median values of $\theta$, and the solid vertical lines with marks
denote the spread of $\theta$ in the distance bin
[$D_{\theta}, D_{\theta}+d D_{\theta}]$. 
The distribution of $\theta$ shows a bi-modal pattern divided around an
intermediate distance at $ D_{\theta}\sim 20$ Mpc \cite{dasetal08}.
The events with small $D_{\theta}$ are likely to be the cases in which
both the source and observer belong to the same filament.
These particles are more likely to travel through strongly magnetized
filaments rather than void regions, resulting in large deflection angles.
On the other hands, for the events with large $D_{\theta}$, the particles
come from distant sources associated with different filaments.
Some of UHECRs may fly through voids and arrive with small $D_{\theta}$,
while most are deflected significantly by the IGMF.
The U-shape distribution of $\langle \theta \rangle$ with minimum at
$ D_{\theta} \sim 20$ Mpc, as well as the U-shape distribution of
$\langle S \rangle$ in Figure 1, are the consequence of the bi-modal
pattern, and a clear signature of the magnetized LSS of the universe. 

\begin{figure}[!t]
\centering
\includegraphics[width=3.2in]{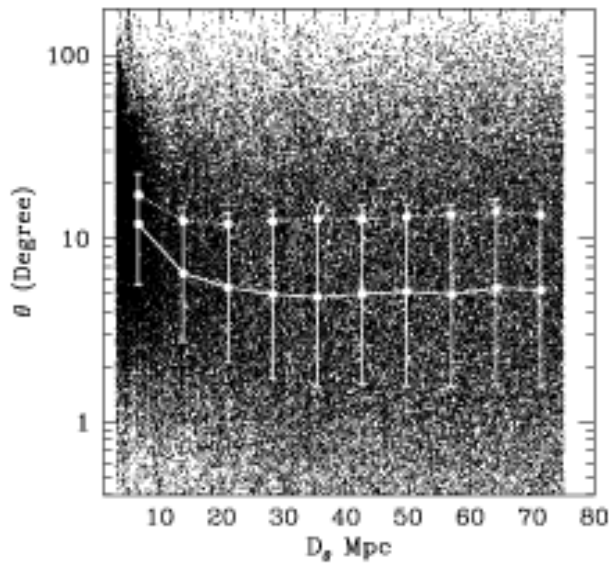}
\caption{Distribution of deflection angle ($\theta$) as functions of
distance ($D_{\theta}$) of true sources. 
The mean (solid line) and median (dotted line) values of ($\theta$)
among the events from source with the distance bin of
$[D_{\theta}, D_{\theta}+d D_{\theta}]$ are shown.
The solid vertical lines with marks connect the first and third quartiles
in the given $D_{\theta}$ bin.}
\end{figure}

The mean deflection angle of all the simulated events is
$\langle \theta \rangle_{\rm sim}$ = $14^{\circ}$, which is $\sim 3$ times
larger than $\langle S \rangle_{\rm sim}$.
One might wonder if deflection angles of this magnitude could erase the
anisotropy in the arrival directions of UHECRs, which is observed in the
Auger data.
However, the large deflection angle does not necessarily lead to the
general isotropy of UHECR arrival directions, since the IGMF distribution
is also correlated with the LSS.
Suppose UHECRs are ejected from sources inside the Local Supercluster.
Some of them will fly along the Supergalactic plane and arrive at Earth with
relatively small $\theta$. 
Some of them may be deflected into void regions, but they may not get
reflected back to the direction toward us due to lack of the turbulent
IGMF there.
In this scenario, the irregularities in the IGMF serve as the `scatters'
of UHECRs.
So we will observe fewer UHECRs from void regions where both sources and
scatters are underpopulated.  
Consequently, the anisotropic behavior in arrival directions could be
maintained even for the large values of $\langle \theta \rangle$,
substantially larger than that of $\langle S \rangle$, if both source
locations and the IGMF distribution are anisotropic in the local universe. 

In order to test the validity of our models for the distributions of the
IGMF and nearby AGNs, we compare the distribution of $S$ obtained from our
simulation with that of the Auger data.
In Figure, 4 we show the cumulative fraction of events, $F(\le \log S)$,
for the Auger data (open circles) and the simulation data (solid line). 
The Kolmogorov-Smirnov (K-S) test yields the maximum difference of $D = 0.17$
between the two distributions, so the significance level of $P\sim 0.37$. 
It means that we cannot reject the null hypothesis that the two distributions
are statistically identical.

\begin{figure}[!t]
\centering
\includegraphics[width=3.2in]{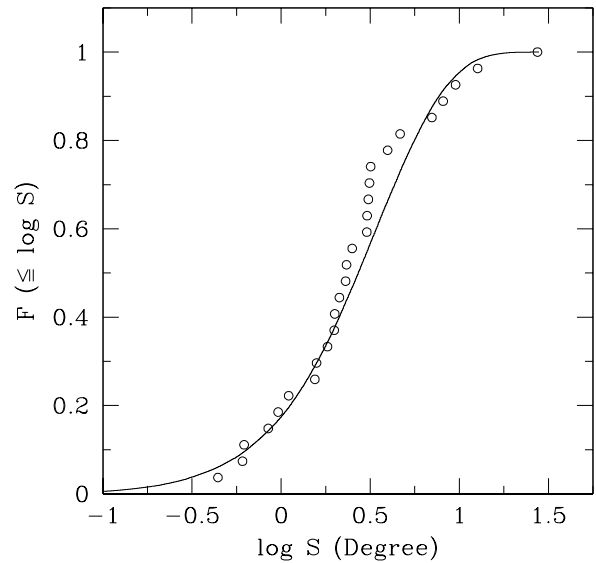}
\caption{Cumulative fraction of events with separation angle smaller than $S$. 
Open circles denote the result for the 27 Auger events \cite{auger08}.
Solid curve is from our simulation. 
The K-S test implies the two distributions are consistent with each other.}
\end{figure}

\begin{figure}[!t]
\centering
\includegraphics[width=3.2in]{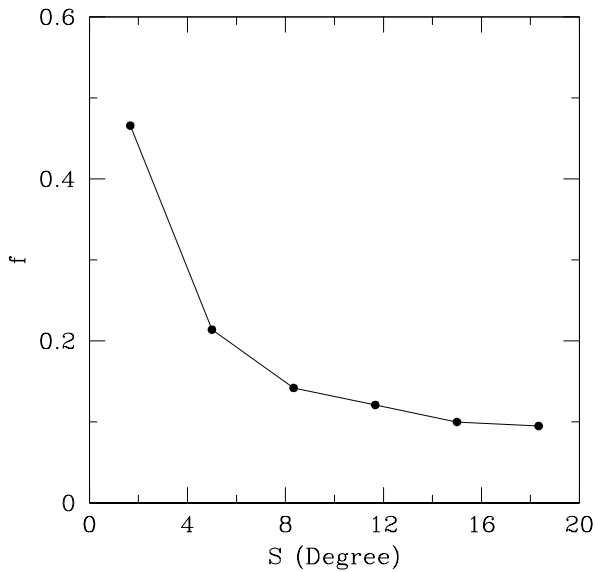}
\caption{Fraction ($f$) of recorded UHE proton events for which their true
sources are identified as the closest AGNs in the sky, as a function
of separation angle ($S$).} 
\end{figure}

\subsection{Probability of Finding UHE Proton Sources}

Since the deflection angle caused by the IGMF is larger than the mean angular
separation, that is, $\langle \theta \rangle > \langle S \rangle$, there
is a good chance that for a given UHECR event, the closest source in the sky
may not be its true source.
To explore the implication, we calculate the fraction of true
identification, $f$, as the ratio of the number of events for which nearest
objects are in fact true sources to the total number of simulated events.
This is a measure of the probability to find the true sources of UHE protons
in our model, when the nearest candidates are chosen blindly (which is the
best we can do with observed data). 
In Figure 5, we show the fraction as function of separation angle, $S$.
For $S \sim 2^{\circ}$, the fraction is $\sim 50$ \%.
As separation angle increases, the fraction decreases gradually to
$\sim 10$ \%, indicating lower probability to find the true sources at
larger separation angles. 
On average, in only 1 out of $\sim 3$ cases, the true sources of UHE protons
can be identified, if we assume that our model for the IGMF is valid. 

\section{Summary}

In the search for the sources of UHECRs, understanding the propagation of
charged particles through the magnetized LSS of the universe is important, 
if UHECRs originate from extragalactic sources. 
At present, the details of the IGMF is still uncertain, mainly due to the
limited available information from observation. 
In this contribution, we adopted a realistic model universe that was
described by a cosmological structure formation simulation \cite{ryuetal08}. 
Our simulated universe represents a characteristic volume of the cosmic
web, in which the distribution of the IGMF was calculated by a physical
model based on turbulent dynamo.
Hypothetical sources of UHE protons were placed at the deepest gravitational
potential wells along the LSS, and observers were placed at groups of
galaxies like our Local Group.

We then followed the propagation of UHE protons above 60 EeV ejected from
the hypothetical sources within a sphere  of radius 75 Mpc with observers
at the center, in the simulated model universe \cite{dasetal08}.
We calculated two angular distances for a given UHE proton event:
the separation angle $S$ from its nearest source candidate and the
deflection angle $\theta$ from it true source.
The distribution of $S$ depends on the relative sky distributions of UHECR
events and source candidates.
We find the mean value of $S$ for simulated events is
$\langle S \rangle_{\rm sim} \approx 3.6^{\circ}$,
while $\langle S \rangle_{\rm Auger} \approx 4.13^{\circ}$ for the 27
Auger events ($\langle S \rangle_{\rm Auger} \approx 3.2^{\circ}$, if
the event with $S > 27^{\circ}$ is excluded).
On the other hand, the distribution of $\theta$ is greatly affected
by the IGMF in the LSS.
The mean value is $\langle \theta \rangle_{\rm sim} \approx 3
\langle S \rangle_{\rm sim} \approx 14^{\circ}$ for the simulated events.
(For the 27 Auger events, of course, there is no way to estimate the
true deflection angles.)

We tested whether the distributions of separation angle, $S$, for
simulated events and the 27 Auger events are identical.
The significance level that the two distributions are drawn from the
identical population is $P\sim 0.37$, according the Kolmogorov-Smirnov test.
Thus, we argue that our simulation results are in a fair agreement with
the Auger data.

We point that, although the mean deflection angle
$\langle \theta \rangle_{\rm sim}$ is large, the arrival directions of
UHECRs would still carry the anisotropy signature of our local universe,
since the IGMF is also correlated with the matter distribution in the LSS.

The fact that $\langle \theta \rangle_{\rm sim} > \langle S \rangle_{\rm sim}$
implies that some of the identified source candidates may not be the
true sources, if we assume that our model of the IGMF is correct.
We estimate the probability of true identification as the ratio of the
number of true source identification to the total number of
simulated events.
This probability is $\sim 50$ \% at $S \sim 2^{\circ}$, but only $20 - 30$
\% for $S=3^{\circ}-4^{\circ}$ and decreases to $\sim 10$ \% at larger
separation angle. 
This suggests that identifying the sources of UHECRs from positional
correlation analysis may not be straightforward.

Finally, we note that $\langle S \rangle$ and $\langle \theta \rangle$
show U-shape distributions, resulting from the bimodal pattern
in which $\theta$ is on average larger for either nearby sources
($D_{\theta} <\ \sim 10$ Mpc) or distant sources ($D_{\theta} >\ \sim 30$ Mpc)
with its minimum at the intermediate distance, $D_{\theta} \sim 20$ Mpc.
This is a characteristic imprint of the structured IGMF to the deflection
angle distribution.
We suggest that the imprint can be tested with the distribution of $S$,
when enough observed UHECR events are accumulated.

The work of DR and HK was supported by the Korea Research Foundation Grant
funded by the Korean Government (MOEHRD) (KRF-2007-341-C00020).

\end{document}